\documentclass[conference]{IEEEtran}
\usepackage{graphicx}
\usepackage{subfig}
\usepackage{color, soul}
\usepackage{cite}
\usepackage{amssymb}
\usepackage{amsmath}
\sethlcolor{green}

\usepackage{algorithm2e}

\makeatletter
\def\ps@headings{%
\def\@oddhead{\mbox{}\scriptsize\rightmark \hfil \thepage}%
\def\@evenhead{\scriptsize\thepage \hfil \leftmark\mbox{}}%
\def\@oddfoot{}%
\def\@evenfoot{}}
\makeatother
\begin{document}

\title{LiteLab: Efficient Large-scale Network Experiments}

\author{\IEEEauthorblockN{Liang Wang, Jussi Kangasharju}\\
\IEEEauthorblockA{Department of Computer
Science, University of Helsinki, Finland}}
%
 

\maketitle
\begin{abstract}
  Large-scale network experiments is a challenging
  problem. Simulations, emulations, and real-world testbeds all have
  their advantages and disadvantages.  In this paper we present
  LiteLab, a light-weight platform specialized for large-scale
  networking experiments. We cover in detail its design, key features,
  and architecture. We also perform an extensive evaluation of
  LiteLab's performance and accuracy and show that it is able to both
  simulate network parameters with high accuracy, and also able to
  scale up to very large networks. LiteLab is flexible, easy to
  deploy, and allows researchers to perform large-scale network
  experiments with a short development cycle.  We have used LiteLab
  for many different kinds of network experiments and are planning to
  make it available for others to use as well.



\end{abstract}


\section{Introduction}
\label{sec:introduction}



For network researchers, large-scale experiment is an important tool
to test distributed systems. A system may exhibit quite different
characteristics in a complex network as opposed to behavior observed
in small-scale experiments. Thorough experimental evaluation before
real-life deployment is very useful in anticipating problems.
%
%
%
This means that the system should be tested with various parameter
values, like different network topologies, bandwidths, link delays,
loss rates and so on. It is also useful to see how the system
interacts with other protocols, e.g. routers with different queueing
policies.

Due to the large parameter space, researchers usually need to run
thousands of experiments with different parameter combinations. As
Eide pointed out in \cite{Eide:2010}, replayability is critical in
modern network experiments.  Not being able to replay an experiment
implies the results are not reproducible, which makes it difficult to
evaluate a system, because the results from different experiments are
not comparable.

A simulator has been a popular option due to its simplicity and
controllability. It also has other benefits like reproducible results
and low resource requirements. However, a simulator is only as good as
the models used. Choosing the right granularity of abstraction is a
tradeoff between more realistic results and increased computational
complexity.


Experiments on real systems can overcome many problems of simulators,
because all the traffic flows through a real network with real-world
behavior. However, running large-scale real-world experiments requires
a lot of resources. Virtualization may help, but configuring and
managing large experiments is still difficult.

Recently, high performance clusters are becoming common,
virtualization technology advances, and overlay networks seem to
become the de facto paradigm for modern distributed systems. All these
emerging technologies change the way we build and evaluate networked
systems.

In this paper, we present \emph{LiteLab}, our flexible platform for
large-scale networking experiments. We show its design,
functionalities, key features and an evaluation of its accuracy and
performance. With LiteLab, researchers can easily construct complex
network on a resource-limited infrastructure.

 
LiteLab is easy to configure and extend. Each router and links between
them can be configured individually. New queueing policies, caching
strategies and other network models can be added in without modifying
existing code.  The flexible design enables LiteLab to simulate both
routers and end systems in the network. Researchers can easily plug in
user application and study system behaviors. LiteLab takes advantages
of overlay network techniques, providing a flexible experiment
platform with many uses.  It helps researchers reduce the experiment
complexity and speeds up experiment life-cycle, and at the same time,
provides satisfying accuracy.

%

The organization of this paper is as follows. Section
\ref{sec:background} discusses choices and existing solutions for
network experiment platforms. Section \ref{sec:architecture} shows the
design of LiteLab and its key features. We evaluate our platform and
also provide some use cases in Section \ref{sec:evaluation}. We
conclude the paper in Section \ref{sec:conclusion}.

\section{Background}
\label{sec:background}

There are generally two methodologies to evaluate a system:
model-based evaluation and experiment-based evaluation. The first
tries to derive numeric performance values by applying analytical
models. However, as the systems become larger and more distributed,
system analysis becomes more complex. In most cases, it is impossible
to build a mathematically tractable model. Experiment-based evaluation
tackles this problem and we can identify three main forms of
experiment-based evaluation: simulation, emulation, and real network
testbeds. Below we present examples of these and discuss their pros
and cons. After presenting LiteLab, we return to a comparison between
LiteLab and the approaches below in Section~\ref{sec:comparison}.



\subsection{Simulator: NS2 and NS3}
\label{sec:background:simulator}
NS2~\cite{NS} is one of the most famous among general purpose
simulators~\cite{NS, disa2:omnet}.  It provides lots of models for
many kinds of network settings. Users can implement their own model in
C++ and plug it into NS2. Experiment configuration and deployment are
done with Tcl/Tk scripts.
NS3 has more features and tunable parameters to allow more realistic
settings. However, increased complexity and detailed models
significantly increase computational overheads.

\subsection{Emulator: Emulab}
Emulab~\cite{White:osdi02} tries to integrate simulation, emulation
and live network into a common framework. The aim is to combine the
control and ease of use from simulation to emulation with the realism
from live network. Users can configure the topology according to the
experiment needs. The multiplexed virtual nodes are implemented with
OpenVZ, a container-based virtualization scheme. Emulab uses VLAN and
Dummynet\cite{Rizzo:1997:DSA:251007.251012} to emulate wide-area links
within the local-area network. It is also able to mix traffic from
Internet, NS3 and emulated links together.  Emulab can also multiplex
NS3 into the experiment in order to maximize the resource utilization.
Emulab's way of combining many advanced techniques makes its
configuration and setup quite complicated. 


\subsection{Internet: PlanetLab}
PlanetLab\cite{Chun:2003:POT:956993.956995,
  Peterson:2003:BID:774763.774772} is a platform for live network
experiments. The generated traffic goes through the real Internet and
is subject to real-life dynamics. PlanetLab takes advantage of its
many geographically distributed sites and provides researchers a
realistic environment very close to the true Internet.  However, as
PlanetLab is a public facility accessible by many researchers, all the
experiments are multiplexed on the same infrastructure. Therefore
experiments are subject to nondeterministic factors, and usually not
repeatable. The experiment configuration also lacks the flexibility of
Emulab.

\section{LiteLab Architecture}
\label{sec:architecture}

We now present the architecture of LiteLab, resource allocation
mechanisms, and mechanisms for running experiments.

\subsection{General Architecture}

\begin{figure}[!tb]
  \centering
  \includegraphics[width=9cm]{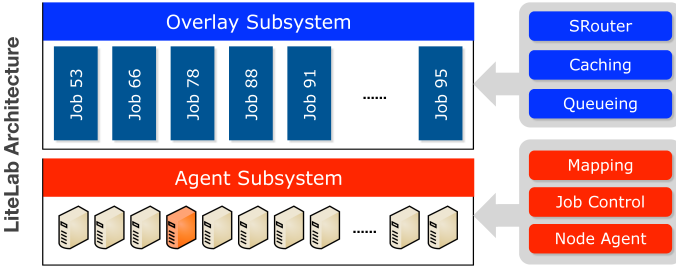}
  \caption{LiteLab Architecture}
  \label{sec:architecture:architecture}
\end{figure}

The goal of LiteLab is to provide an easy to use, fully-fledged
network experiment platform. Figure
\ref{sec:architecture:architecture} shows the general system
architecture. LiteLab consists of two subsystems: \textit{Agent
  Subsystem} and \textit{Overlay Subsystem}, presented
below.
 
We first illustrate how LiteLab works by describing how an experiment
is performed on this experiment platform.

All experiments are \textit{jobs} in LiteLab and are defined by a job
description archive provided by a user. An archive can contain
multiple configuration files which specify the details of the
experiment, e.g., network topology, router configuration, link
properties, etc. The Agent Subsystem has one leader node which is
responsible for starting and managing jobs (see
Section~\ref{sec:architecture:allocation:mapping}). We use the Bully
election algorithm for selecting the leader dynamically.



Second, the user submits the job to LiteLab which processes the job
description archive, determines needed resources and allocates
necessary physical nodes from the available nodes. We have developed
and run LiteLab on our department's cluster, but the design puts no
constraints on where the nodes are located.\footnote{For
  geographically dispersed nodes, strong guarantees about network
  performance may be hard or impossible to provide.} Nodes with
lighter loads are preferred.

Third, LiteLab informs the selected nodes and deploys an instance of
the Overlay Subsystem on them (see below). The Overlay Subsystem is
started to construct the network specified in the job description
archive.

Finally, LiteLab starts the experiment, and the job is saved into the
\emph{JobControl module}, which continuously monitors its state. If a
node is overloaded, LiteLab will migrate some SRouters to other
available nodes to reduce the load. If an experiment successfully
finished, all the log files are automatically collected for post
processing.

\subsection{Agent Subsystem}
\label{sec:architecture:agent}

The Agent Subsystem provides a stable and uniform experiment
infrastructure. It hides the communication complexity, resource
failures and other underlying details from the Overlay Subsystem.  It
is responsible for managing physical nodes, allocating resources,
administrating jobs, monitoring experiments and collecting
results. The main components of the Agent Subsystem are NodeAgent,
JobControl and Mapping.

\begin{enumerate}

\item \textbf{NodeAgent} represents a physical node, thus there is
  one-to-one mapping between the two. It has two major roles in
  LiteLab. First, it serves as the communication layer of the whole
  platform. Second, it presents itself as a reliable resource pool to
  Overlay Subsystem. We use the Bully algorithm to elect a leader
  responsible for managing the resources and jobs.

 

\item \textbf{JobControl} manages all the submitted jobs in LiteLab.
  After pre-processing, JobControl allocates the resources and splits
  a job into multiple tasks which are distributed to the selected
  nodes. The job is started and continuously monitored.

\item \textbf{Mapping} maps virtual resources to physical
  resources. The goal is to maximize resource utilization and perform
  the mapping quickly. It is also a key component to guarantee the
  accuracy. Mapping module runs an LP solver to achieve the goal.

\end{enumerate}


\subsection{Resource Allocation: Static Mapping}
\label{sec:architecture:allocation:mapping}

Resource allocation focuses on the mapping between virtual nodes and
physical nodes, and it is the key to platform scalability.
We have subdivided the resource allocation problem into two
sub-problems: \textit{mapping problem} (below) and \textit{dynamic
  migration} (Section~\ref{sec:architecture:allocation:migration}).

The mapping should not only maximize the resource utilization, but
also guarantee there is no violation of physical capacity. We take
four metrics into account as the constraints: CPU load, network
traffic, memory usage and use of pseudo-terminal devices.  Deployment
of the software-simulated routers (SRouter) must respect the physical
constraints while optimize the use of physical resources.

Suppose we have $m$ physical nodes and $n$ virtual nodes.  We first
construct an $m \times n$ deployment matrix $D$. All the elements in
$D$ have binary values.
If $D_{i,j}$ is $1$, then virtual node $i$ is deployed on physical
node $j$, otherwise $D_{i,j}$ is 0. We denote $C_i$ as the CPU power,
$M_i$ as the memory capacity, $U$ as egress bandwidth and $V$ as
ingress bandwidth of physical node $i$. We also denote $c_j$, $m_j$,
$u_j$ and $v_j$ as virtual node $j$'s requirements for CPU, memory,
egress and ingress bandwidth respectively.


We model the processing capability of a node in terms of its CPU
power:

\begin{equation}
  \label{eq:first}
  \sum_{j=1}^{n} D_{i,j} \times c_j \le C_i, \forall i \in \{1, 2, 3 .. m\}
\end{equation}

Total memory requirements from virtual nodes running on the same
machine should not exceed its physical memory:

\begin{equation}
  \sum_{j=1}^{n} D_{i,j} \times m_j \le M_i, \forall i \in \{1, 2, 3 .. m\}
\end{equation}

The aggregated bandwidths are also subject to physical node's
bandwidth limit:

\begin{equation}
  \sum_{j=1}^{n} D_{i,j} \times u_j \le U_i, \forall i \in \{1, 2, 3 .. m\}
\end{equation}

\begin{equation}
  \sum_{j=1}^{n} D_{i,j} \times v_j \le V_i, \forall i \in \{1, 2, 3 .. m\}
\end{equation}

A virtual node can only be deployed on one physical node, and the
total number of virtual nodes is fixed. Two natural constraints
follow:

\begin{equation}
  \sum_{i=1}^{m} D_{i,j} = 1, \forall j \in \{1, 2, 3 .. n\}
\end{equation}

\begin{equation}
  \label{eq:last}
  \sum_{i=1}^{m} \sum_{j=1}^{n} D_{i,j} = n
\end{equation}

Our mapping strategy is to use as few physical nodes as possible, and
give preference to less loaded nodes. In other words, we try to deploy
as many virtual nodes as possible on physical nodes with the lightest
load. We define node load $L$:

\begin{equation}
  \label{eq:lp:load}
  \begin{split}
    L = w_1 \times {avg\_cpu\_load} + w_2 \times traffic \\
    + w_3 \times memory\_usage + w_4 \times user\_activities
  \end{split}
\end{equation}

The four metrics are given different weights to reflect different
level of importance. In our case, we set $w_1 > w_2 > w_3 > w_4$, but
this choice is rather arbitrary; Emulab uses a similar
rationale~\cite{White:osdi02}. In our evaluation, we have found that
our simple rule for the weights is sufficient, but further study would
be required to gain more understanding on their importance.



Larger $L$ implies heavier load.  We give each machine a preference
factor $p_i$ equal to the reciprocal of its load, $L^{-1}$. Node with
the lightest load has the largest preference index.

We formalize the mapping problem into a \textit{linear programming
  problem} (LP). The objective function is as follows:

\begin{equation}
  \label{eq:lp:objective}
  \max \sum_{i=1}^{m} \sum_{j=1}^{n}{p_i \times D_{i,j}}
\end{equation}
subject to the constraints in
equations~\eqref{eq:first}--\eqref{eq:last}.

\begin{figure}[!tb]
  \centering
  \includegraphics[width=9cm]{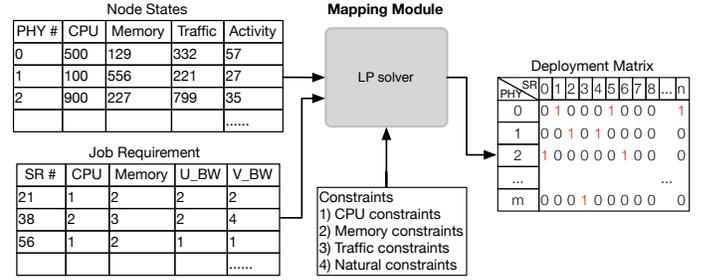}
  \caption{Inputs and output of Mapping module}
  \label{sec:architecture:mapping}
\end{figure}

Each node sends its state information to the leader, which then has
global knowledge needed for solving the LP problem. Our LP solver is a
python module, which takes node states and job description as inputs,
and outputs the optimal deployment matrix. Figure
\ref{sec:architecture:mapping} shows how Mapping module works.

We also adopted other mechanisms into our LP solver to further improve
the efficiency by reducing the problem complexity. We discuss these in
detail in Section \ref{sec:evaluation:smart}.


\subsection{Resource Allocation: Dynamic Migration}
\label{sec:architecture:allocation:migration}

The static mapping cannot efficiently handle the dynamics during an
experiment.  For example, a node overloaded by other users' activities
may skew our experiment results. We use dynamic migration to solve
these problems.

Dynamic migration is implemented as a sub-module in NodeAgent. It
keeps monitoring the load (defined by e.q (\ref{eq:lp:load})) on its
host.  If NodeAgent detects a node is overloaded, some tasks will be
moved onto other machines without restarting the
experiments. Migration is not able to completely mask the effects from
other users, but can alleviate the worst problems. Currently, we only
implement very basic migration. Thorough testing and more features are
part of our future work.

\subsection{Overlay Subsystem}
\label{sec:architecture:overlay}

\begin{figure}[!tb]
  \centering
  \includegraphics[width=9cm]{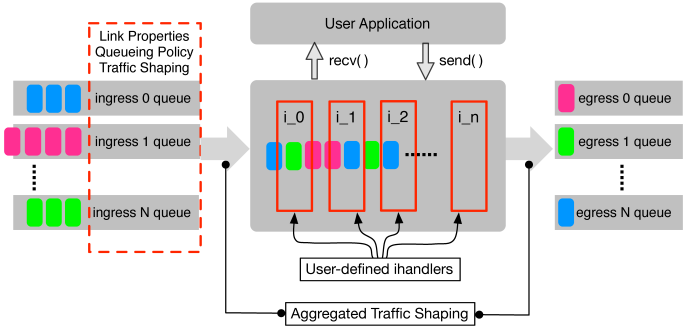}
  \caption{Logical structure of SRouter}
  \label{sec:architecture:srouter_architecture}
  \vskip -5mm
\end{figure}

Overlay Subsystem constructs an experiment overlay by using the
resources from Agent Subsystem. One overlay instance corresponds to a
job, therefore LiteLab can have multiple overlay instances running in
parallel at the same time. JobControl module manages all the created
overlays.

The most critical component in Overlay Subsystem is SRouter, which is
a software abstraction of a realistic router. Due to its light-weight,
multiple SRouters can run on one physical node. Users can configure
many parameters of SRouters, e.g., link properties (delay, loss rate,
bandwidth), queue size, queueing policy(Droptail, RED), and so
on.

\subsubsection{Queues}
Figure \ref{sec:architecture:srouter_architecture} shows SRouter
architecture and how the packets are processed inside. SRouter
maintains a TCP connection to each of its neighbors. The connection
represents a physical link in real-world. For each link, SRouter
maintains a queue to buffer incoming packets. In the job description
archive, user can specify the link delay, loss rate and bandwidth for
each individual link. All these properties are modelled within the
SRouter so that they are not subject to TCP dynamics.

SRouter maintains three FIFO queues inside: \textit{iqueue},
\textit{equeue} and \textit{cqueue}. All incoming packets are pushed
into \textit{iqueue} before being processed by a chain of
functions. All outgoing packets are pushed into
\textit{equeue}. Aggregated traffic shaping is done on these two
queues.  If a packet's destination is the current SRouter itself, then
it enters into \textit{cqueue}.  Later, the packet will be delivered
to user application.

\subsubsection{Processing Chain}

We borrowed the concept of chains of rules from \textit{iptables} when
we designed SRouter. If a packet waiting in the ingress queue gets its
chance to be processed, it will go through a chain of functions, each
of which can modify the packet. We call such a function an
\textit{ihandler}. If an ihandler decides to drop the packet, then the
packet will not be passed to the rest of the ihandlers in the chain.

The default ihandler is \textit{bypass\_handler}, and is always the
last one in the chain. It simply inserts an incoming packet into
\textit{cqueue} or \textit{equeue}. If a packet reaches the last
ihandler, it will be either delivered to the next hop or to the user
application.

Users can insert their own ihandlers into SRouter to process incoming
packets. SRouter has a very simple but powerful mechanism to load
user-defined ihandlers. User only needs to specify the path of the
folder containing all the ihandlers in job description archive. After
LiteLab starts a job, SRouter will load them one by one, in the
specified order. 

\subsubsection{VID}

To be neutral to any naming scheme, LiteLab uses logical ID (VID) to
identify a SRouter. A VID can be an integer, a float number or an
arbitrary string. By using VID, we do not need to allocate separate IP
address for each SRouter. Every VID is mapped to $<$IP:PORT$>$ tuple
and an SRouter maintains a table of such mappings. These mappings are
also a key to enabling dynamic migration, because LiteLab can migrate
an SRouter onto another node by simply updating the VID mapping table.

\subsubsection{Routing}
Routing in LiteLab is also based on VID. In LiteLab, an experimenter
can use his own routing algorithm either by plugging in ihandlers or
by defining a static routing policy.
LiteLab has three default routing mechanisms:

\begin{enumerate}

\item \textbf{OTF}: Uses OSPF\cite{RFC2328} protocol. Given the
  topology file as input, the routing table is calculated on the fly
  when an experiment starts. The routing table construction is fast,
  but the routes are not symmetric.

\item \textbf{SYM}: Symmetric route is needed in some experiments. In
  such cases, LiteLab uses \textit{Floyd-Warshall
    algorithm}\cite{Floyd:1962:A9A:367766.368166} to construct routing
  table. In the worst case, \textit{Floyd-Warshall} has time
  complexity of $O(|V|^3)$ and space complexity of $\Theta(|V|^2)$.
  Therefore, the construction time might be long if the topology is
  extremely complex.

\item \textbf{STC}: This method loads the routing table from a file,
  avoiding the computational overheads in the other two methods and
  giving full control over routing.

\end{enumerate}

\subsubsection{User Application}
ihandler provides a passive way to interact with SRouters.  Besides
ihandler, SRouter has another mechanism for users to interact with it:
\textit{user application}. This feature makes it possible to use
SRouter as a functional end-system. In the beginning of an experiment,
LiteLab will also start the user applications running on SRouters
after they successfully load all the ihandlers.

SRouter exposes two interfaces to user application: \textit{send} and
\textit{recv}. Equivalent to standard socket calls, a user application
can use them sending and receiving packets. Currently, we have a
synchronous version implemented, an asynchronous version is in our
future work. User applications can also access various other
information like VID, routing table, link usage, etc.

In a nutshell, LiteLab is highly customizable and extensible. It is
very simple to plug in user-defined modules without modifying the code
and substitute default modules.
SRouter can also be used as end-system instead of simply doing routing
task. When used as end-system, user-implemented applications can be
run on top of
it. 

\section{Evaluation}
\label{sec:evaluation}



We performed thorough evaluation on LiteLab to test its accuracy,
performance and flexibility. In following sections, we present how we
evaluates various aspects of LiteLab and how we adapted it to get
around practical issues. We also give some use cases to illustrate the
power of the platform.

\subsection{Accuracy: Link Property}
\label{sec:evaluation:validate}

In terms of accuracy, we have two concerns with using
software-simulated router on general purpose operating system.  First,
is SRouter able to saturate the emulated link if it is operating at
full speed? Second, can SRouter emulate the link properties (delay,
loss rate and bandwidth) accurately?

We performed a series of experiments to test the accuracy and
precision of SRouter. We used different values for bandwidth, delay,
packet loss, and packet size in the evaluation. Our test nodes run
Ubuntu SMP with kernel 3.0.0-15-server. The operating system clock
interrupt frequency is 250HZ.
We set up an experiment where we use two SRouters as end-systems, each
running both a server and a client.

In bandwidth limit experiments, we used one-way traffic. A server
keeps sending packets to a client on the other
node. Table~\ref{sec:evaluation:tab:bandwidth} summarizes the
experiment results.  With 1518~B packets, SRouter can easily saturate
a 100~Mbps link. With 64~B packets, two directly connected SRouters
can exchange 10000 packets (625~KB) per second. This low number stems
mainly from our use of Python to implement LiteLab. Although C would
be faster, we opted for Python in the interest of simplicity and
flexibility.
We also tested a multi-hop scenario and observed only negligible
additional decreases in bandwidth.

Compared with the results in \cite{White:osdi02}, SRouter is much
better than~\textit{nse}~\cite{780820} and close to dummynet. One
reason why dummynet has slightly better accuracy is it increases the
clock interrupt frequency of the emulation node to 10000HZ, 40 times
of ours, which improves the precision accordingly. Another reason is
that dummynet works at the kernel level thus has no user-level
overheads. Based on the results, SRouter makes a reasonable tradeoff
between accuracy and complexity. It shows that application layer
isolation is able to provide satisfying accuracy and precision.  We
can increase experiment scale greatly without sacrificing too much
reality.



Table \ref{sec:evaluation:tab:delay} summarizes the results from delay
test.  We used the same topology as in the bandwidth limit test, with
the difference that traffic is two-way and there is no bandwidth
limit.
In an ideal situation, the observed value should be twice the set
value. The results show that as the delay increases, the error drops
even though the standard deviation (stdev) also increases. Both small
and large packets suffer from large error rate when the delay is less
than 10ms.
We also noticed that a high-speed network (10~Gbps) can provide better
precision than a low-speed network (1~Gbps) in both experiments and
comparing with previous work\cite{White:osdi02}.

Table \ref{sec:evaluation:tab:lossrate} summarizes the experiments for
packet loss observed by the customer. The accuracy of the modeled loss
rate mainly relies on the pseudo-random generator in the language.

\begin{table}[!tb]
  \caption{Accuracy of SRouter's bandwidth control as a function of link bandwidth and packet size.}
  \begin{tabular}{ | p{1.2cm} | p{1cm} || p{2.2cm} | p{2.2cm} | }
    \hline
    Bandwidth & Packet & \multicolumn{2}{c|}{Observed Value} \\
    (Kbps)    & Size   & bw (Kbps) & \% err \\
    \hline
    56 & 64 & 55.77 & 0.41 \\
    & 1518 & 57.62 & 2.89 \\
    \hline
    384 & 64 & 382.56 & 0.37 \\
    & 1518 & 387.96 & 1.03 \\
    \hline
    1544 & 64 & 1539.23 & 0.31 \\
    & 1518 & 1546.32 & 0.15 \\
    \hline
    10000 & 1518 & 9988 & 0.12 \\
    \hline
    45000 & 1518 & 44947 & 0.12 \\
    \hline
  \end{tabular}
  \label{sec:evaluation:tab:bandwidth}
  \vskip -5mm
\end{table}

\begin{table}[!tb]
  \caption{Accuracy of SRouter's delay at maximum packet rate as a
    function of 1-way link delay and packet size.}
  \begin{tabular}{ | p{1.2cm} | p{1cm} || p{1.33cm} | p{1.33cm} | p{1.33cm} | }
    \hline
    OW Delay & Packet & \multicolumn{3}{c|}{Observed Value} \\
    (ms)   & Size   & RTT & stdev & \% err\\
    \hline
    0 & 64 & 0.190 & 0.004 & N/A \\
    & 1518 & 0.221 & 0.007 & N/A \\
    \hline
    5 & 64 & 10.200 & 0.035 & 2.00 \\
    & 1518 & 10.230 & 0.009 & 2.30 \\
    \hline
    10 & 64 & 20.212 & 0.057 & 1.06\\
    & 1518 & 20.185 & 0.015& 0.92 \\
    \hline
    50 & 64 & 100.209 & 0.060 & 0.21 \\
    & 1518 & 100.218 & 0.031 & 0.22 \\
    \hline
    300 & 64 & 600.189 & 0.083 & 0.03 \\
    & 1518 & 600.273 & 0.034 & 0.04 \\
    \hline
  \end{tabular}
  \label{sec:evaluation:tab:delay}
\end{table}

\begin{table}[!tb]
  \caption{Accuracy of SRouter's packet loss rate as a function of link loss rate and packet size.}
  \begin{tabular}{ | p{1.2cm} | p{1cm} || p{2.2cm} | p{2.2cm} | }
    \hline
    Loss Rate & Pakcet & \multicolumn{2}{c|}{Observed Value} \\
    (\%)    & Size   & loss rate (\%) & \% err \\
    \hline
    0.8 & 64 & 0.802 & 0.2 \\
    & 1518 & 0.798 & 0.2 \\
    \hline
    2.5 & 64 & 2.51 & 0.4 \\
    & 1518 & 2.52 & 0.8 \\
    \hline
    12 & 64 & 11.98 & 0.1 \\
    & 1518 & 11.97 & 0.2 \\
    \hline
  \end{tabular}
  \label{sec:evaluation:tab:lossrate}
\end{table}

\subsection{Scalability: Topology}
\label{sec:evaluation:topology}

Being able to quickly construct new topologies certainly improves
efficiency.  Compared to Emulab and PlanetLab, LiteLab's configuration
and setup is lighter. Nodes are identified with VIDs and no additional
IP addresses are needed.


\begin{table}[!tb]
  \caption{Time to construct realistic ISP networks. OTF: routing
    table is calculated on the fly; STC: routing table is pre-computed
    and loaded by routers} 
  \begin{tabular}{ | p{1.2cm} | p{1.3cm} | p{1.3cm} | p{1.3cm} | p{1.3cm} | }
    \hline
    ISP & \# of routers & \# of links & OTF & STC \\
    \hline
    Exodus & 248 & 483 & 1.5s & 16s \\
    \hline
    Sprint & 604 & 2279 & 4.6s & 141s \\
    \hline
    AT\&T & 671 & 2118 & 4.2s & 204s \\
    \hline
    NTT & 972 & 2839 & 10.1s &  312s\\
    \hline
  \end{tabular}
  \label{sec:evaluation:tab:topology}
  \vskip -5mm
\end{table}

To test how fast LiteLab can construct an experiment network, we used
both synthetic and realistic topologies, deployed on 10 machines. For
realistic topologies, we used 4 ISPs' router-level networks from
Rocketfuel Project~\cite{SpringN:Rocketfuel}.
Table~\ref{sec:evaluation:tab:topology} shows the time used in
constructing these networks using two different routing table
computing methods (OTF and STC from
Section~\ref{sec:architecture:overlay}).  The result shows LiteLab is
very fast in constructing realistic networks, most of them finished
within 5 seconds. Even for the largest network NTT, the time to
construct is only about 10 seconds.

In some cases, the experimenters need symmetric routes. As we
mentioned in Section~\ref{sec:architecture}, network constructed with
OTF cannot guarantee symmetric routes. SYM is impractical for complex
topologies, so STC is the only option, although much slower than
OTF. The first bottleneck is transmitting the pre-computed routing
file to the local machine; the second bottleneck is loading the
routing table into the memory. There are several ways to speed up STC:
first, storing the routing file in local file system; second, using
more physical nodes.

\begin{figure}[!tb]
  \centering \subfloat[Random
  (Erd\H{o}s-R\'{e}nyi)]{\label{sec:evaluation:fig:topology:1}\includegraphics[width=4.2cm]{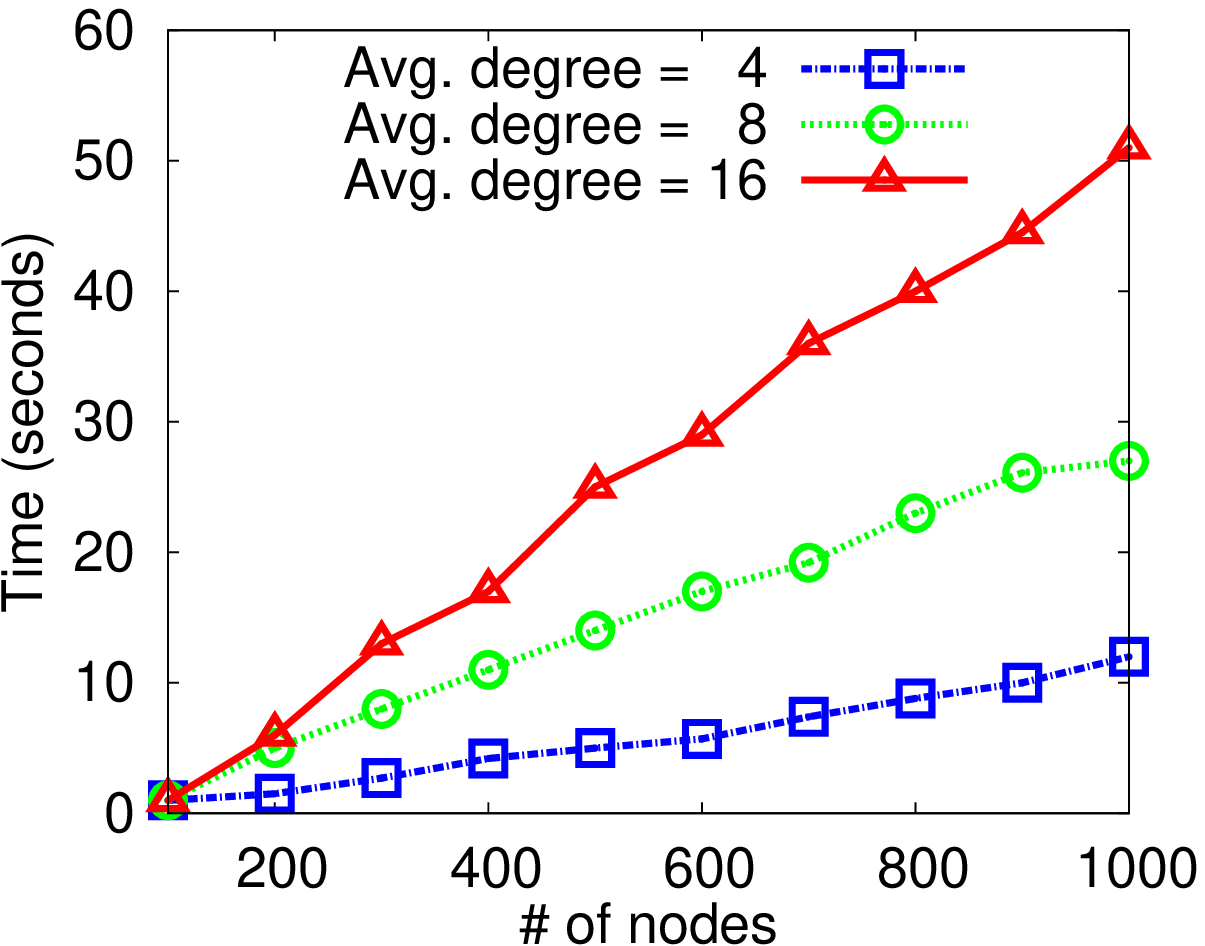}}
  \quad \subfloat[Scale-free
  (Barab\'{a}si-Albert)]{\label{sec:evaluation:fig:topology:2}\includegraphics[width=4.2cm]{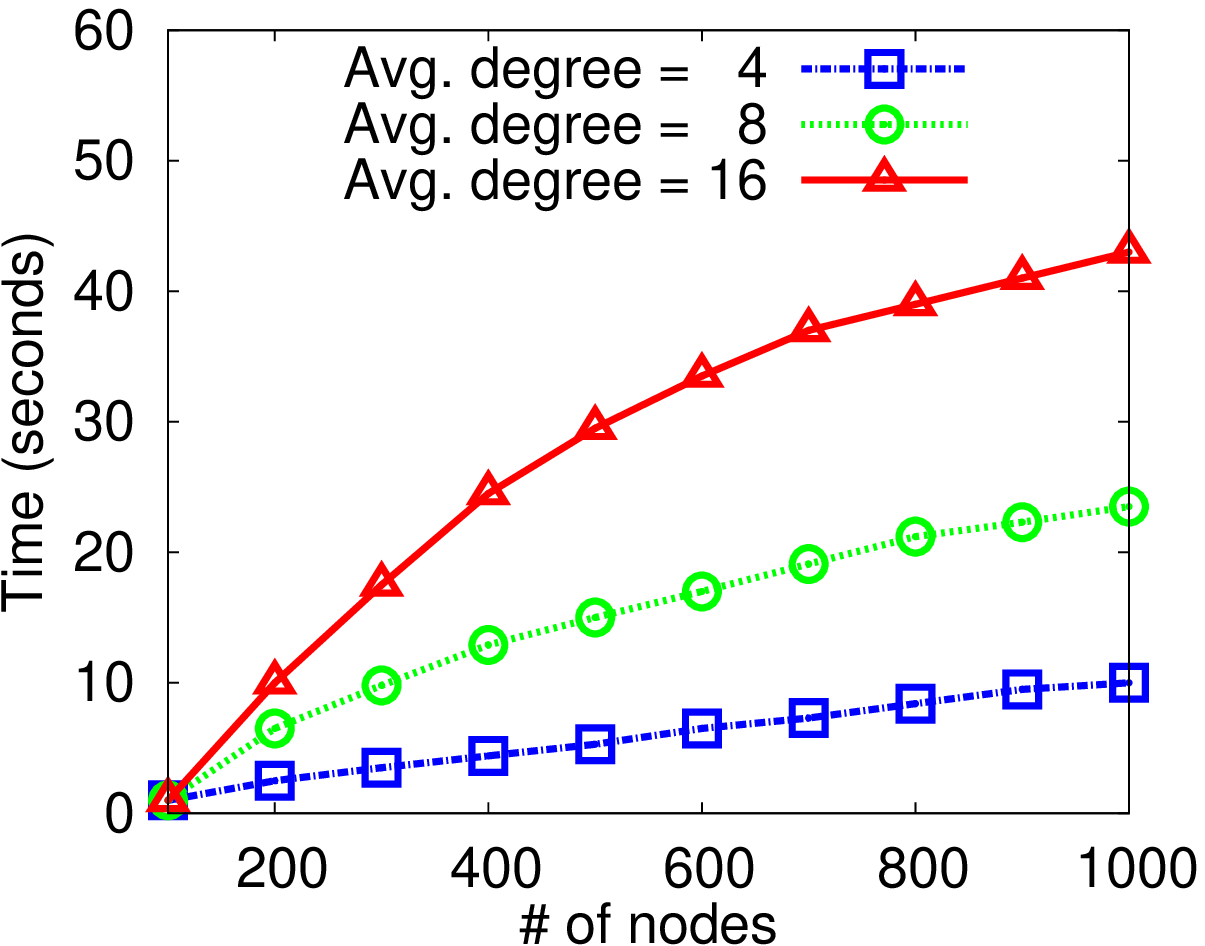}}
  \caption{Time to construct synthetic networks of different type.}
  \label{sec:evaluation:fig:topology}
  \vskip -2mm
\end{figure}

We also used synthetic network topologies in the evaluation. The
purpose is to illustrate the relation between construction time and
network complexity. We chose Erd\H{o}s-R\'{e}nyi model to generate
random network and Barab\'{a}si-Albert model to generate scale-free
network.  Figure \ref{sec:evaluation:fig:topology} shows the results
of 50 different synthetic networks with the different average node
degrees. The number of nodes increases from 100 to 1000. In the
biggest network, there are about 16000 links.  From the results, we
can see given the node degree, the time to construct network increases
linearly as the number of nodes increases in random network. However,
the growth of time is slower in scale-free network because the nodes
with high degree dominate the construction time.


%

\subsection{Adaptability: Resources Allocation}
\label{sec:evaluation:smart}

As mentioned, we use an LP solver to map virtual nodes to physical
nodes.
The LP solver module uses
\textit{CBC}\footnote{https://projects.coin-or.org/Cbc} as engine,
takes node states and job requirements as inputs, and outputs a
deployment matrix.
We tested how well our LP solver scales by giving it different size of
inputs. Table~\ref{sec:evaluation:tab:lpsolver} shows the results.

\begin{table}[!tb]
  \begin{tabular}{ | p{1.6cm} | p{1.6cm} | p{1.6cm} | p{1.6cm} | }
    \hline
    \# of PHY & \# of SR & Naive (s) & Heur (s) \\
    \hline
    128 & 100 & 1.30 & 0.03 \\
    \hline
    128 & 200 & 3.23 & 0.08 \\
    \hline
    128 & 400 & 5.37 & 0.23 \\
    \hline
    128 & 800 & 11.61 & 1.00 \\
    \hline
    \hline
    256 & 100 & 1.93 & 0.03 \\
    \hline
    256 & 200 & 4.62 & 0.08 \\
    \hline
    256 & 400 & 9.47  & 0.23 \\
    \hline
    256 & 800 & 22.24 & 0.98 \\
    \hline
  \end{tabular}
  \caption{Efficiency of LP solver as a function of different network size. \textbf{PHY}:physical nodes, \textbf{SR}:SRouters}
  \label{sec:evaluation:tab:lpsolver}
  \vskip -5mm
\end{table}

The size of deployment matrix is the product of the number of physical
nodes and the number of SRouters.  The results (\textit{Naive} column
in Table \ref{sec:evaluation:tab:lpsolver}) show that as the
deployment matrix grows, the solving time increases.  It implies that
even for a moderate overlay network, solving times can be significant
if there are a lot of physical resources.

\begin{figure}[!tb]
  \centering
  \includegraphics[width=8cm]{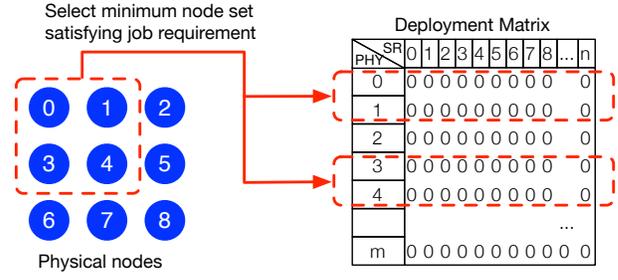}
  \caption{Reduce deployment matrix size by selecting minimum physical
    node set that satisfies the job requirement.}
  \label{sec:evaluation:fig:mininodeset}
\end{figure}

To reduce solving time, we must reduce matrix size. We cannot reduce
the number of SRouters in the experiment, but can limit the number of
physical nodes. Algorithm~\ref{alg:mapping:heur} shows our heuristic
algorithm which attempts to limit the number of physical nodes. The
algorithm picks the minimum number of nodes needed to satisfy the
\emph{aggregate resource requirements} of the experiment and then
attempts to solve the LP. Because the actual requirements like CPU or
memory of a single SRouter cannot be split onto two physical nodes, it
is possible that the LP has no solution. We then double the aggregate
resources required, add more physical nodes, and attempt to solve the
LP again. Eventually a solution will be found or the problem will be
deemed infeasible, i.e., although the aggregate resources are
sufficient, it is not possible to find a mapping which satisfies
individual node and SRouter constraints. In our tests, we discovered
that the optimal solution is in most cases found on the first
try. Figure \ref{sec:evaluation:fig:mininodeset} shows how the matrix
size is reduced.



\begin{algorithm}[!tb]
  \caption{Heuristic to improve mapping efficiency}
  \label{alg:mapping:heur}
  {\textbf{Input:} job requirements \textbf{R}, physical nodes \textbf{L}} \\
  {\textbf{Output:} minimum physical node set \textbf{S}} \\
  Calculate overall job requirement \textbf{R} \\
  Order nodes from lightest to heaviest load into \textbf{L} \\
  \ForEach{ node \textbf{N} in \textbf{L} } {
    Add \textbf{N} to \textbf{S} \\
    Calculate capacity of \textbf{S}: \textbf{C} \\
    \If{ \textbf{R} $<$ \textbf{C} } {
      Solve LP \\
      \lIf{ optimal solution exists } {
        break \\
      } \lElse { $\textbf{R} \leftarrow 2 \times \textbf{R}$ } } }
\end{algorithm}

The efficiency of the LP solver with heuristic algorithm is also shown
in Table \ref{sec:evaluation:tab:lpsolver} (Heur column). By comparing
with naive LP solver, the solving time is significantly reduced and
\textit{it is independent of the number of available physical nodes}.





\subsection{Application: Use Cases}
\label{sec:evaluation:usecase}

We have used LiteLab as an experiment platform in many
projects. Compared to other platforms, LiteLab speeds up experiment
life-cycle without sacrificing accuracy, especially for very complex
experiment networks. We have tested LiteLab in the following
situations:

\begin{enumerate}

\item Router experiment: new queueing and caching algorithms can be
  plugged into LiteLab and test its performance with various link
  properties.

\item Distributed algorithms: LiteLab is also a good platform for
  studying distributed algorithms, gossip and various DHT protocols
  can be implemented as user applications.

\item Information-centric network experiments: various routing and
  caching algorithms can be easily tested on complex networks, using
  realistic ISP networks.
\end{enumerate}




\subsection{Limitations}
\label{sec:evaluation:limitation}

LiteLab aims at being a flexible, easy-to-deploy experiment platform
and in this goal, it must make some tradeoffs regarding accuracy and
performance. In terms of accuracy, the main factor is the precision of
the system interrupt timer, especially when simulating low-level link
properties. Increasing timer frequency, like in~\cite{White:osdi02},
will improve accuracy, but requires root privileges and possible
recompilation of the kernel. LiteLab runs in user space and does not
require root privileges.


Overall system load will also affect LiteLab's performance. LiteLab
attempts to avoid these issues by selecting lightly loaded nodes and
migrating tasks from heavily loaded nodes, however it cannot
completely eliminate external effects from other processes running on
the test platform. Any platform on a shared infrastructure suffers
from this same problem and the only solution would be to use a
dedicated infrastructure.


SRouter's processing power is another limitation as it can only
process about 10000 packets per second. Adding more user-defined
modules will further slow down SRouter. This limitation stems mainly
from our choice of Python as implementation language and a C
implementation would yield better performance. Using Python has made
the development of LiteLab a lot easier and less error-prone than
using C. This means that LiteLab is not well-suited for low-level,
fine-grained protocol work. However, for studying system-level
behavior and performance of a large-scale system, it is better suited
than existing platforms.



\subsection{Comparison}
\label{sec:comparison}


We now compare features and capabilities of LiteLab with the three
other existing approaches. LiteLab is a time- and space-shared
experiment platform. It leverages the existing nodes available to the
experimenters and attempts to maximize utilization of all available
physical resources.

Compared to NS2/3 (and other simulators like
\cite{disa2:omnet,ccnsim,ndnsim}), LiteLab runs over a real network
and allows deployment of user applications on top of the experiment
platform. LiteLab allows experimenting with very large topologies with
relatively little physical resources. Specific-purpose simulators,
e.g.,~\cite{PeerSim, CPE:CPE710, 4301435, YangW:GPS}, are lighter than
NS2/3, but are limited to a single application. Parallel
simulation~\cite{Fujimoto:1989:PDE:76738.76741} may offer a solution
to the scalability issues of simulators.



Compared with Emulab, LiteLab runs on generic hardware and does not
require any particular operating system or root privileges. Emulab is
more accurate in simulating certain network-level parameters, but
LiteLab is able to run a much larger experiment with the same hardware
because multiple SRouters can run on a single physical node.  Work of
Rao et al.~\cite{5569970} is close to our approach. However, their
work focuses on a specific application whereas LiteLab is a generic
network experimentation testbed.

LiteLab is very similar to PlanetLab, with a few key
differences. PlanetLab runs on a dedicated infrastructure whereas
LiteLab can leverage any existing infrastructure. PlanetLab has the
advantage of using a real network between the nodes, but at the same
time, is not able to guarantee network performance between
nodes. LiteLab, on the other hand, can configure the network
properties with very good accuracy and allow better repeatability for
experiments.

\section{Conclusion}
\label{sec:conclusion}

In this paper, we have presented LiteLab, a light-weight network
experiment platform.  It combines the benefits from both emulation and
simulation: ease of use, high accuracy, no complicated hardware
settings, easy to extend and interface with user application, various
operating parameters to reflect realistic settings.

LiteLab is a flexible and versatile platform to study system behavior
in complex networks. It makes large-scale experiments possible even
with limited physical resources. LiteLab also shortens experiment
life-cycle without sacrificing the realism, makes researcher's work
more efficient. We are making LiteLab available for download and use
of others in the future.





\end{document}